\documentclass[ALICE,manyauthors]{cernphprep}
\usepackage[comma,square,numbers,sort&compress]{natbib}

\usepackage{multirow}
\usepackage{lineno}
\usepackage{xspace}
\usepackage{hyperref}
\usepackage{color}

\usepackage{upgreek}

\usepackage[T1]{fontenc}
\usepackage{orcidlink}

\begin{document}
%

\newcommand{\pp}           {pp\xspace}
\newcommand{\ppbar}        {\mbox{$\mathrm {p\overline{p}}$}\xspace}
\newcommand{\XeXe}         {\mbox{Xe--Xe}\xspace}
\newcommand{\PbPb}         {\mbox{Pb--Pb}\xspace}
\newcommand{\pA}           {\mbox{pA}\xspace}
\newcommand{\pPb}          {\mbox{p--Pb}\xspace}
\newcommand{\AuAu}         {\mbox{Au--Au}\xspace}
\newcommand{\dAu}          {\mbox{d--Au}\xspace}

\newcommand{\s}            {\ensuremath{\sqrt{s}}\xspace}
\newcommand{\snn}          {\ensuremath{\sqrt{s_{\mathrm{NN}}}}\xspace}
\newcommand{\pt}           {\ensuremath{p_{\rm T}}\xspace}
\newcommand{\meanpt}       {$\langle p_{\mathrm{T}}\rangle$\xspace}
\newcommand{\ycms}         {\ensuremath{y_{\rm CMS}}\xspace}
\newcommand{\ylab}         {\ensuremath{y_{\rm lab}}\xspace}
\newcommand{\etarange}[1]  {\mbox{$\left | \eta \right |~<~#1$}}
\newcommand{\yrange}[1]    {\mbox{$\left | y \right |~<~#1$}}
\newcommand{\dndy}         {\ensuremath{\mathrm{d}N_\mathrm{ch}/\mathrm{d}y}\xspace}
\newcommand{\dndeta}       {\ensuremath{\mathrm{d}N_\mathrm{ch}/\mathrm{d}\eta}\xspace}
\newcommand{\avdndeta}     {\ensuremath{\langle\dndeta\rangle}\xspace}
\newcommand{\dNdy}         {\ensuremath{\mathrm{d}N_\mathrm{ch}/\mathrm{d}y}\xspace}
\newcommand{\Npart}        {\ensuremath{N_\mathrm{part}}\xspace}
\newcommand{\Ncoll}        {\ensuremath{N_\mathrm{coll}}\xspace}
\newcommand{\dEdx}         {\ensuremath{\textrm{d}E/\textrm{d}x}\xspace}
\newcommand{\RpPb}         {\ensuremath{R_{\rm pPb}}\xspace}

\newcommand{\nineH}        {$\sqrt{s}~=~0.9$~Te\kern-.1emV\xspace}
\newcommand{\seven}        {$\sqrt{s}~=~7$~Te\kern-.1emV\xspace}
\newcommand{\twoH}         {$\sqrt{s}~=~0.2$~Te\kern-.1emV\xspace}
\newcommand{\twosevensix}  {$\sqrt{s}~=~2.76$~Te\kern-.1emV\xspace}
\newcommand{\five}         {$\sqrt{s}~=~5.02$~Te\kern-.1emV\xspace}
\newcommand{\twosevensixnn}{$\sqrt{s_{\mathrm{NN}}}~=~2.76$~Te\kern-.1emV\xspace}
\newcommand{\fivenn}       {$\sqrt{s_{\mathrm{NN}}}~=~5.02$~Te\kern-.1emV\xspace}
\newcommand{\LT}           {L{\'e}vy-Tsallis\xspace}
\newcommand{\GeVc}         {Ge\kern-.1emV/$c$\xspace}
\newcommand{\GeVcc}        {Ge\kern-.1emV/$c^{2}$\xspace}
\newcommand{\MeVc}         {Me\kern-.1emV/$c$\xspace}
\newcommand{\TeV}          {Te\kern-.1emV\xspace}
\newcommand{\GeV}          {Ge\kern-.1emV\xspace}
\newcommand{\MeV}          {Me\kern-.1emV\xspace}
\newcommand{\GeVmass}      {Ge\kern-.2emV/$c^2$\xspace}
\newcommand{\MeVmass}      {Me\kern-.2emV/$c^2$\xspace}
\newcommand{\lumi}         {\ensuremath{\mathcal{L}}\xspace}

\newcommand{\ITS}          {\rm{ITS}\xspace}
\newcommand{\TOF}          {\rm{TOF}\xspace}
\newcommand{\ZDC}          {\rm{ZDC}\xspace}
\newcommand{\ZDCs}         {\rm{ZDCs}\xspace}
\newcommand{\ZNA}          {\rm{ZNA}\xspace}
\newcommand{\ZNC}          {\rm{ZNC}\xspace}
\newcommand{\SPD}          {\rm{SPD}\xspace}
\newcommand{\SDD}          {\rm{SDD}\xspace}
\newcommand{\SSD}          {\rm{SSD}\xspace}
\newcommand{\TPC}          {\rm{TPC}\xspace}
\newcommand{\TRD}          {\rm{TRD}\xspace}
\newcommand{\VZERO}        {\rm{V0}\xspace}
\newcommand{\VZEROA}       {\rm{V0A}\xspace}
\newcommand{\VZEROC}       {\rm{V0C}\xspace}
\newcommand{\Vdecay} 	   {\ensuremath{V^{0}}\xspace}

\newcommand{\ee}           {\ensuremath{e^{+}e^{-}}} 
\newcommand{\pip}          {\ensuremath{\pi^{+}}\xspace}
\newcommand{\pim}          {\ensuremath{\pi^{-}}\xspace}
\newcommand{\kap}          {\ensuremath{\rm{K}^{+}}\xspace}
\newcommand{\kam}          {\ensuremath{\rm{K}^{-}}\xspace}
\newcommand{\pbar}         {\ensuremath{\rm\overline{p}}\xspace}
\newcommand{\kzero}        {\ensuremath{{\rm K}^{0}_{\rm{S}}}\xspace}
\newcommand{\lmb}          {\ensuremath{\Lambda}\xspace}
\newcommand{\almb}         {\ensuremath{\overline{\Lambda}}\xspace}
\newcommand{\Om}           {\ensuremath{\Omega^-}\xspace}
\newcommand{\Mo}           {\ensuremath{\overline{\Omega}^+}\xspace}
\newcommand{\X}            {\ensuremath{\Xi^-}\xspace}
\newcommand{\Ix}           {\ensuremath{\overline{\Xi}^+}\xspace}
\newcommand{\Xis}          {\ensuremath{\Xi^{\pm}}\xspace}
\newcommand{\Oms}          {\ensuremath{\Omega^{\pm}}\xspace}
\newcommand{\degree}       {\ensuremath{^{\rm o}}\xspace}

\newcommand{\pSig}{\ensuremath{\mathrm {p\mbox{--}\Sigma^{+}}}\,}
\newcommand{\aprot}{\ensuremath{\mathrm{\bar{p}}}\,}
\newcommand{\prot}{\ensuremath{\mathrm{p}}\,}
\newcommand{\pP}{\ensuremath{\mathrm {p\mbox{--}p}}\,}
\newcommand{\pXi}{\ensuremath{\mathrm {p\mbox{--}\Xi}}\,}
\newcommand{\pOmega}{\ensuremath{\mathrm {p\mbox{--}\Omega}}\,}
\newcommand{\LXi}{\ensuremath{\mathrm {\Lambda\mbox{--}\Xi}}\,}
\newcommand{\phiP}{\ensuremath{\mathrm {\phi\mbox{--}p}}\,}
\newcommand{\akP}{\ensuremath{\mathrm {K^-\mbox{--}p}}\,}
\newcommand{\Kp}{\ensuremath{\mathrm {K\mbox{--}p}}\,}
\newcommand{\aKp}{\ensuremath{\mathrm {K^+\mbox{--}p}}\,}
\newcommand{\Kap}{\ensuremath{\mathrm {K^-\mbox{--}\overline{p}}}\,}
\newcommand{\pipi}{\ensuremath{\mathrm {\uppi^+\mbox{--}\uppi^+}}\,}
\newcommand{\apiapi}{\ensuremath{\mathrm {\uppi^-\mbox{--}\uppi^-}}\,}
\newcommand{\spipi}{\ensuremath{\mbox{$\uppi$--$\uppi$}}~}
\newcommand{\pap}{\ensuremath{\mathrm {p\mbox{--}\bar{p}}}\,}
\newcommand{\paL}{\ensuremath{\mathrm {p\mbox{--}\overline{\Lambda}}}\,}
\newcommand{\apL}{\ensuremath{\mathrm {\overline{p}\mbox{--}\Lambda}}\,}
\newcommand{\LaL}{\ensuremath{\mathrm {\Lambda\mbox{--} \overline{\Lambda}}}\,}
\newcommand{\pL}{\ensuremath{\mathrm {p\mbox{--}\Lambda}}\,}
\newcommand{\ApaL}{\ensuremath{\mathrm {\bar{p}\mbox{--}\overline{\Lambda}}}\,}
\newcommand{\BBbar}{\ensuremath{\mathrm {B\mbox{--} \bar{B}}}\,}
\newcommand{\BB}{\ensuremath{\mathrm {B\mbox{--} B}}\,}
\newcommand{\rhop}{\ensuremath{\mathrm {\rho^{0}\mbox{--} p}}\,}
\newcommand{\rhoap}{\ensuremath{\mathrm {\rho^{0}\mbox{--} \overline{p}}}\,}

\newcommand{\Ledn}         {Lednick\'y--Lyuboshits\xspace}
\newcommand{\rs}           {\ensuremath{r^*}\xspace}
\newcommand{\rsv}           {\ensuremath{\vec{r}^*}\xspace}
\newcommand{\rc}           {\ensuremath{r_\mathrm{core}}\xspace}
\newcommand{\rcv}           {\ensuremath{\vec{r}_\mathrm{core}}\xspace}
\newcommand{\rcs}           {\ensuremath{r^*_\mathrm{core}}\xspace}
\newcommand{\rcsv}           {\ensuremath{\vec{r}^*_\mathrm{core}}\xspace}
\newcommand{\ks}           {\ensuremath{k^*}\xspace}
\newcommand{\ksv}           {\ensuremath{\vec{k}^*}\xspace}
\newcommand{\Sr}           {\ensuremath{S(r)}\xspace}
\newcommand{\Ck}           {\ensuremath{C(k)}\xspace}
\newcommand{\Srs}           {\ensuremath{S(\rs)}\xspace}
\newcommand{\Cks}           {\ensuremath{C(\ks)}\xspace}
\newcommand{\mt}           {\ensuremath{m_{\mathrm{T}}}\xspace}

\newcommand{\dedx}{d$E$/d$x$}
\newcommand{\dndydpt}{${\rm d}^2N/({\rm d}y {\rm d}p_{\rm T})$}
\newcommand{\pbpb}{Pb--Pb}
\newcommand{\sig}{$\Sigma^{+}$}
\newcommand{\asig}{$\overline{\Sigma}^{-}$}
\newcommand{\sigo}{$\Sigma^{0}$}

\begin{titlepage}
\PHyear{2025}       
\PHnumber{218}      
\PHdate{22 September}  

\title{Measurement of the p\mbox{--}$\Sigma^+$ correlation function in pp collisions at \boldmath{$\sqrt{\textit{s}}=13$}~TeV}
\ShortTitle{p\mbox{--}$\Sigma^+$ correlation function in pp collisions at \boldmath{$\sqrt{\textit{s}}=13$}~TeV}   

\Collaboration{ALICE Collaboration\thanks{See Appendix~\ref{app:collab} for the list of collaboration members}}
\ShortAuthor{ALICE Collaboration} 

\begin{abstract}
In this letter, the first measurement of the femtoscopic correlation of protons and \sig\ hyperons is presented and used to study the p\mbox{--}\sig\ interaction. The measurement is performed with the ALICE detector in high-multiplicity triggered pp collisions at $\sqrt{s} = 13$~TeV. The \sig\ hyperons are reconstructed using a missing-mass approach in the decay channel to $\textrm{p} + \uppi^0$ with $\uppi^0\rightarrow\upgamma\upgamma$, while both \sig\ and protons are identified using a machine learning approach. These techniques result in a high reconstruction efficiency and purity, which allows the measurement of the p\mbox{--}\sig\ correlation function for the first time. Thanks to the high significance achieved in the p\mbox{--}\sig\ correlation signal, it is possible to discriminate between the predictions of different models of the N\mbox{--}$\Sigma$ interaction and to accomplish a first determination of the p\mbox{--}\sig\ scattering parameters.

\end{abstract}

\end{titlepage}

\setcounter{page}{2} 


\section{Introduction} 

Quantitative knowledge of the hyperon\mbox{--}nucleon interaction provides an important contribution to a comprehensive understanding of the role of strangeness in quantum chromodynamics\,\cite{Botta:2012xi,Gal:2016boi,Ratti:2018ksb,Tolos:2020aln,Adolfsson:2020dhm,Harris:2023tti}. Furthermore, it is a crucial ingredient for the equation of state of hadronic matter at densities that are two or three times higher than the nuclear saturation density. Such conditions may occur, for instance, in the interior of neutron stars\,\cite{Schaffner-Bielich:2010csv}. At sufficiently high nuclear densities, the Pauli principle requires the nucleons to occupy such high energy states that it eventually becomes energetically favorable for hyperons, mainly the $\Lambda~(J^P = \frac{1}{2}^+)$ isospin singlet ($I=0$, $S=-1$), to occur. This leads to a softening of the equation of state, i.e. the pressure stabilizing the star against gravity increases slower with increasing density and hence less mass can be supported. This is, however, incompatible with the observation that heavy neutron stars may reach up to two solar masses. Recently, it was pointed out that this so-called hyperon puzzle could be resolved by a repulsive N\mbox{--}N\mbox{--}$\Lambda$ three-body force that may inhibit the occurrence of $\Lambda$ baryons in neutron stars\,\cite{ALICE:2022boj, Gerstung:2020ktv, Vidana:2024ngv, Lonardoni:2014bwa}. Nonetheless, other hyperons, in particular 
the $\Sigma^-$, might still be present\,\cite{Baldo:1998hd,Vidana:2024ngv}.
Unfortunately, there is no experimental information on the interaction of $\Sigma^-$ with neutrons\,\cite{Tan:1973at}.
However, one can alternatively consider p\mbox{--}\sig\ as a guideline, since it is expected to be practically identical to the former (disregarding the Coulomb interaction), given that the effects from charge-symmetry breaking are presumably small. 
Nonetheless, data on the p\mbox{--}\sig\ interaction remain scarce; as a result, theoretical models are poorly constrained, and substantial discrepancies between available calculations persist. Particularly, interaction in the triplet channel ($^3\textrm{S}_1$) of the N\mbox{--}$\Sigma$ system
with isospin $I=3/2$ is not well-known, and the predictions in the 
literature vary from attraction to repulsion\,\cite{Rijken:1998yy,Nagels:1973rq,Haidenbauer:2005zh,Fujiwara:2006yh,
Nagels:2015lfa,Haidenbauer:2019boi,Haidenbauer:2021zvr}.
The extension of the femtoscopic studies of the ALICE Collaboration to the $\Sigma$ sector can improve our current experimental situation by providing valuable information on $\Sigma$ interactions with nucleons.

In a pioneering study, the ALICE Collaboration investigated the p\mbox{--}\sigo\ interaction by the measurement of the correlation function of protons and \sigo\,\cite{ALICE:2019buq}. The p\mbox{--}\sigo\ correlation function is an important input for the measurement of the p\mbox{--}$\Lambda$ correlation function, as they cannot be experimentally separated. While this analysis proved the principal feasibility of the measurement using the femtoscopy technique, which complements pertinent experimental efforts at J-PARC\,\cite{J-PARCE40:2022nvq} to access the strong interaction in the $\Sigma$ sector, experimental challenges hindered a definite conclusion on the N\mbox{--}$\Sigma$ strong interaction. In particular, insufficient number of events and low purity led to a low significance of the signal.

Driven by advances in the reconstruction technique of low-energy photons, ALICE recently delivered the first measurement of the \sig\ baryon production at the LHC\,\cite{ALICE:2025hqt}. This paves the way for further investigations of the N\mbox{--}$\Sigma$ interaction using femtoscopy. Exploiting these techniques to the fullest, this letter presents the first measurement of the p\mbox{--}\sig\ correlation function. Thanks to the high significance that can be achieved, it is possible for the first time to discriminate between different model predictions and to deduce first results for the p\mbox{--}\sig\ scattering parameters from the measurement.

\section{Data analysis}
\label{sec:Data_analysis}

A detailed description of the ALICE detector setup and its performance in LHC Runs 1 and 2 can be found in Ref.~\cite{Aamodt:2008zz,ALICE:2014sbx,ALICE:2022wpn}. In the following, only a brief description of the subdetectors used in this analysis is given. Those are the V0 detectors\,\cite{ALICEVZERO}, the Inner Tracking System (ITS)\,\cite{ALICEITS}, the Time Projection Chamber (TPC)\,\cite{ALICETPC}, and the Time-Of-Flight detector (TOF)\,\cite{ALICETOF}.

The V0 detectors consist of two plastic scintillator arrays located at forward $(2.8 < \eta < 5.1)$ and backward $(-3.7 < \eta < -1.7)$ pseudorapidities and are used for event triggering. If the signal amplitude in the V0 detectors exceeds a certain threshold, the event is classified as a high-multiplicity (HM) one. The ITS is a six-layer cylindrical silicon detector used for tracking, vertexing, and triggering. The layers are located at radii between 3.9~cm and 43~cm around the beam axis. The inner two layers of the ITS are also used for triggering. The TPC is a cylindrical gaseous detector located around the ITS with an inner radius of around 85~cm and an outer radius of 250~cm. The TPC is the main tracking detector and also contributes to the determination of the primary vertex. Furthermore, the TPC is used for particle identification (PID) through the measurement of the specific energy loss \dedx\ in the detector gas. Both ITS and TPC are located inside a 0.5~T solenoidal magnetic field and cover a pseudorapidity range of $|\eta| < 0.9$ in the full azimuth. The TOF detector surrounds the TPC and complements the PID through the measurement of the velocity $\beta$ of the particles.

The data analyzed in this letter were recorded between 2016 and 2018 during the LHC pp run at $\sqrt{s} = 13$~TeV. The analysis was performed using a high-multiplicity (HM) triggered sample. In total about $1.3\times 10^9$ HM-triggered events have been analyzed, which corresponds to an integrated luminosity of 14.1~pb$^{-1}$\,\cite{ALICE:2023udb}. The reason for using a high-multiplicity triggered sample is the strongly increased number of particle pairs in these events. The average charged particle multiplicity in $|\eta|<0.5$ ($\langle dN_{ch}/d\eta \rangle$) amounts to $30.8\pm0.4$ in this sample, in contrast to $6.9\pm0.1$ in the minimum-bias sample\,\cite{ALICE:2020swj}. The number of p\mbox{--}\sig\ pairs is further promoted by the increased production of strangeness at high multiplicities.
For the determination of the purity of the particle samples and the resolution of the reconstruction, Monte Carlo (MC) simulations employing the PYTHIA 8 event generator with the Monash 2013\,\cite{Bierlich:2022pfr, Skands:2014pea} tune are used. The simulated particles are propagated through the detectors using GEANT 4\,\cite{Agostinelli:2002hh, Allison:2006ve, Allison:2016lfl} and processed using the ALICE reconstruction algorithm\,\cite{Aamodt:2008zz}.

In the following, ``\sig'' will be used to refer to \sig\ and its anti-particle \asig and the tuple ``p\mbox{--}\sig'' will refer to p\mbox{--}\sig as well as $\bar{\text{p}}$\mbox{--}\asig.

The \sig\ is reconstructed in its hadronic decay channel $\Sigma^{+} \rightarrow \rm{p} + \uppi^{0}$, with the subsequent electromagnetic decay of $\uppi^{0} \rightarrow \upgamma + \upgamma$. Photons can convert into an electron\mbox{--}positron pair within the detector material. The reconstruction of photons from such conversions is called photon conversion method (PCM)\,\cite{ALICE:2017ryd}. The converted photons are reconstructed from pairs of $\textrm{e}^{+}$ and $\textrm{e}^{-}$ tracks during the tracking and selected by their distinct V-shaped topology (V$^0$). The conversion probability of the photons is low (5--8\%) and their reconstruction efficiency vanishes towards low momenta, limiting the reconstruction efficiency of \sig. 

\begin{figure}[t]
\centering
\includegraphics[width=0.49\textwidth]{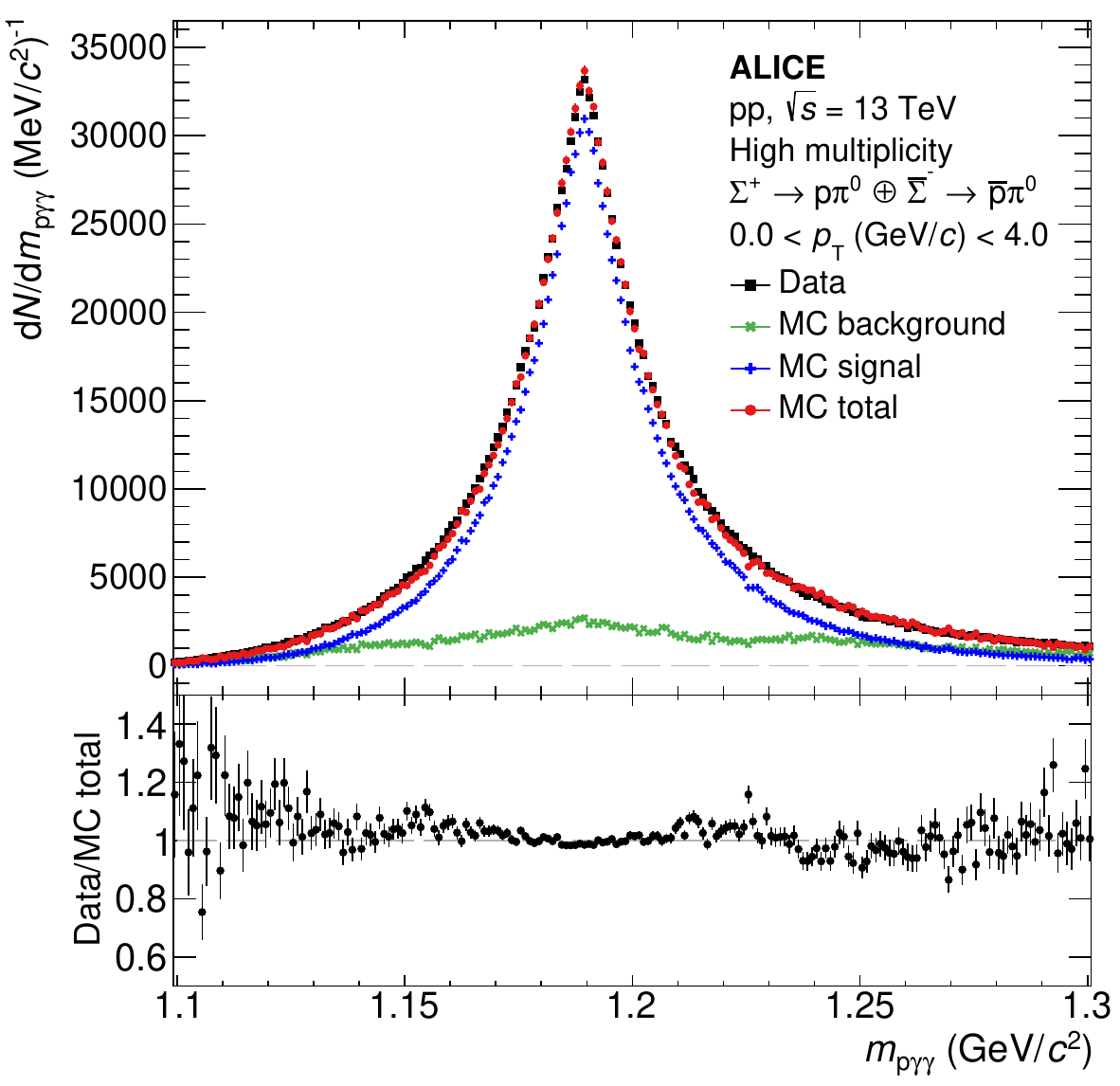}
\includegraphics[width=0.49\textwidth]{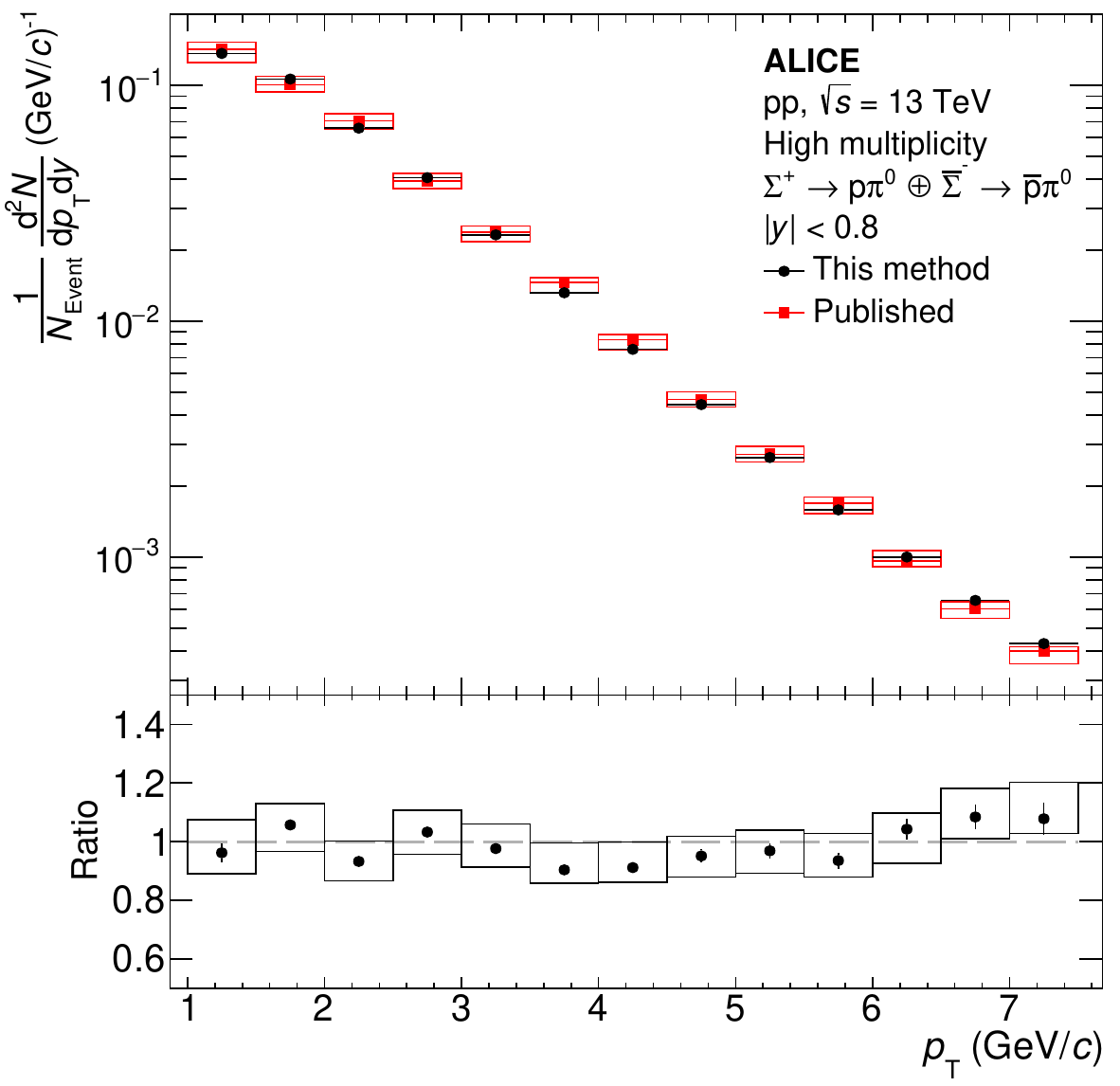}
\caption{Left panel: Invariant-mass distribution of \sig\ candidates in 0.0 $< p_{\rm T} <$ 4.0~$\rm GeV/\it{c}$ (black markers) with MC template fits: signal (blue markers), background (green markers), and total (red markers). The used missing-mass reconstruction method leads to a non-Gaussian shape of the distribution. The ratio of the data and the MC total distribution is shown in the lower panel.\\Right panel: Comparison of the \pt\ spectrum of \sig\ using the reconstruction method introduced in this paper with the corresponding spectrum measured in Ref.~\cite{ALICE:2025hqt}. The ratio of the spectra is shown in the lower panel. The spectra are in good agreement, indicating that the reconstruction method and the purity determination work well over the full \pt\ range.}
\label{fig:invmass_comparison}
\end{figure}

Instead, an alternative approach is pursued in this study, exploiting the unique decay topology in conjunction with momentum conservation, improving the reconstruction efficiency of \sig\ by around one order of magnitude. Only one of the photons is reconstructed via PCM, giving access to its conversion vertex and momentum components. The decay vertex of the \sig\ (secondary vertex) is reconstructed from the PCM photon and the proton using the Kalman Filter package (KFParticle)\,\cite{KFParticle}. Given the negligible lifetime of the intermediate $\uppi^0$, it can be assumed that the proton and the photon originate from a common vertex.\\When the direction of flight of the \sig\ is known from the topology, the momenta of the decay daughters with respect to this direction must cancel. The missing transverse momentum $\vec{p}^{\upgamma2}_{\textrm{T},\Sigma^+}$ of the unobserved photon with respect to the direction of flight of the \sig\ is computed as
\begin{align}
\label{eq:reco1}
\vec{p}^{\upgamma2}_{\textrm{T},\Sigma^+}=-(\vec{p}^{\textrm{\,p}}_{\textrm{T},\Sigma^+}+\vec{p}^{\upgamma1}_{\textrm{T},\Sigma^+})
\end{align}
where $\vec{p}^{\textrm{\,p}}_{\textrm{T},\Sigma^+}$ and $\vec{p}^{\upgamma1}_{\textrm{T},\Sigma^+}$ are the transverse momenta of the proton and the observed photon, respectively, and all kinematic quantities are evaluated with respect to the direction of flight of the \sig. The momentum components of the photons, which point out of the decay plane of the \sig\ (``p$\times$\sig'') need to be carried by the photons alone, such that
\begin{align}
\vec{p}^{\upgamma2}_{\textrm{p}\times \Sigma^+}=-\vec{p}^{\upgamma1}_{\textrm{p}\times \Sigma^+}.
\end{align}
The third momentum component of the unobserved photon can be derived by making use of the known $\uppi^{0}$ mass $m_{\uppi^0}$. To this end, 
\begin{align}
\label{eq:reco2}
m_{\uppi^0} = 2\cdot (|p_{\upgamma1}|\cdot |p_{\upgamma2}|-|p_{\upgamma1}\cdot p_{\upgamma2}|)
\end{align}
is solved for the missing momentum component and the solution minimizing the deviation from the nominal \sig\ mass is selected. This method ensures minimal background impact and minimizes the inclusion of misreconstructed particles in the signal region. The invariant-mass distribution is shown in Fig.~\ref{fig:invmass_comparison} (left panel). As seen in the figure, the missing-mass reconstruction method leads to a non-Gaussian shape of the distribution and a comparably large peak width.

The resulting momentum resolution is dominated by the positional uncertainty of the secondary vertex, with smaller contributions from the momentum resolutions of the proton and the photon. The method is highly sensitive to the secondary-vertex uncertainty. This behavior is expected as Eqs.~\eqref{eq:reco1} and \eqref{eq:reco2} depend on the direction of flight of the \sig, which is calculated from the secondary vertex. In contrast, the momentum resolution of the measured particles is sufficiently small to be neglected. In the following, the secondary-vertex position is considered in spherical coordinates (with polar angle $\theta$ and azimuthal angle $\varphi$). This is practical, as the direction of flight of the \sig\ is fully described by $\theta$ and $\varphi$. The method is insensitive to the radius of the vertex and hence the large uncertainty associated with it. The uncertainties of $\theta$ and $\varphi$ are obtained from the MC simulation. The uncertainty in the secondary vertex both affects the calculated invariant mass and the momentum of the \sig. The coordinates $\theta$ and $\varphi$ of the secondary vertex are varied randomly within 0.02~rad, which corresponds to covering roughly two standard deviations. Thereby, the variations are drawn from their distributions. The values which bring the invariant mass of the \sig\ closest to the nominal mass are used. This leads to an improvement of the momentum resolution, and almost completely recovers the smearing induced by the missing-mass reconstruction. 

The observable of interest is the relative momentum $(\ks~=~\frac{1}{2}|\Vec{p}_{\textrm{\,p}}^*~-~\Vec{p}_{\Sigma^+}^*|)$ of pairs of protons and \sig, evaluated in their center-of-mass frame (denoted with an asterisk (*)). The resolution of \ks is evaluated using the MC simulation as the width of the Gaussian parametrization of the deviation of the reconstructed \ks from the true value in slices of \ks. The resolution amounts to roughly 1\% at high \ks ($>600$~\MeVc) and saturates at 6~\MeVc\ in the relevant region (\ks $<200$~\MeVc). The resolution is well below the bin width of the correlation function (40~\MeVc), particularly in the region below 500 \MeVc, which is the region of interest.

Both \sig\ and protons are identified using boosted decision trees (BDT) using the XGBoost library\,\cite{Chen:2016btl}. For the interface to the ROOT trees Hipe4ML\,\cite{Hipe4ML} is used. The hyperparameter optimization is performed using Optuna\,\cite{Optuna}. For the \sig\ two classes are used (signal, background), while for the protons three classes (primaries, secondaries, misidentified) are used. For the misidentified protons no distinction between primaries and secondaries is made. All training samples (signal and background) are taken from the MC simulation. This is particularly important for the \sig, as the broad peak does not allow the usage of sidebands. The size of the signal and background samples used for the training are balanced. Given the large number of protons in the full data set, only events with at least one selected \sig\ candidate are considered for the proton selection.

In the following, the training parameters for the BDTs will be discussed. For protons, TPC and TOF PID information, the distance-of-closest-approach~(DCA) to the primary vertex (in $xy$ and $z$), the number of clusters in the ITS and the TPC, and the total momentum are used. The PID information is presented in a parameterized way and in absolute terms ($\beta$, \dedx\ for the TOF and TPC, respectively).

For the \sig, the parameters of the proton daughter are equal to those used for the primary protons. For the photons, the opening angle, invariant mass, conversion radius, the absolute momentum and the Armenteros\mbox{--}Podolanski\,\cite{Armenteros} parameters are used. The latter are the longitudinal momentum asymmetry $\alpha=(p_l^+-p_l^-)$/$(p_l^++p_l^-)$ of the daughter tracks and the transverse momentum $q_\text{T}$ of the daughters with respect to the flight direction of the photon. Additionally, the TPC PID information and the number of clusters of the daughter electrons are provided. For the reconstructed \sig, additional parameters are computed, which are the DCA between the proton and the photon, the proper pointing angle of the photon with respect to the decay vertex of the \sig, the flight distance of the \sig, the transverse momentum of the \sig, and the $\chi^2$ of the KFParticle\,\cite{KFParticle} reconstruction.
The hyperparameters are tuned to optimize the performance and minimize negative effects as overtraining. The cut-off in the BDT score is chosen to maximize the purity while preserving the particle yields. This way, a momentum-dependent purity between 75\% and 90\% is achieved for the \sig.
The efficiency-corrected \pt\ spectrum of the \sig, reconstructed and selected with the introduced methods, is compared to the \pt spectrum of \sig, which was measured in a previous analysis in Ref.~\cite{ALICE:2025hqt}. The spectra are in very good agreement as shown in Fig.~\ref{fig:invmass_comparison} (right panel), validating the reconstruction and selection procedure. 
After selecting the protons and \sig, around $2.7\cdot10^6$ protons and $1.8\cdot10^6$ \sig\ are available for further analysis. The raw mean \pt\ of the protons is about 1.4~\GeVc, while that of the \sig\ is significantly larger, as the spectrum is folded with the momentum-dependent reconstruction efficiency, and amounts to about 2.7~\GeVc.

\subsection{Correlation function}

The experimental two-particle correlation function is defined as in Refs.~\cite{Lisa:2005dd, ALICE:2018ysd}
\begin{align}\Cks=\mathcal{N}\times\frac{N_{\mathrm{same}}(\ks)}{N_{\mathrm{mixed}}(\ks)}
\end{align}
where $N_{\mathrm{same}}(\ks)$ and $N_{\mathrm{mixed}}(\ks)$ are the distributions
of the relative momentum \ks in the same and mixed events, respectively. The latter is the \ks distribution of uncorrelated pairs, estimated using the mixed-event technique. The normalization constant is denoted with $\mathcal{N}$ and ensures the proper asymptotic behavior: for large relative momenta \Cks should converge to unity. The measured correlation function $\Cks$ contains both the genuine p\mbox{--}\sig\ correlation function, as well as contributions stemming from all other possible origins (i.e. feed-down, misidentification, and physical background sources), which need to be accounted for. The fractions of the contributions are incorporated in the so-called $\lambda$-parameters, such that the measured correlation function can be written as in Ref.~\cite{ALICE:2018ysd} 
\begin{equation}
    \Cks= \sum_i \lambda_i C_i(\ks)
\end{equation}
where the index $i$ runs over all contributions to the measured correlation function. The values of the $\lambda$-parameters are determined in a data-driven way from single-particle properties, and can be estimated by the product of the purity and primary fractions. 
For the protons, misidentified particles and secondary protons (from weak decays of $\Lambda$ and \sig) are considered, which results in a total of six contributions. Judging from previous analyses, knock-out (i.e. particles knocked out of the detector material) is negligible in proton\mbox{--}proton collisions\,\cite{ALICE:2018ysd}. For the \sig, only the combinatorial background under the peak needs to be considered, as there is no feed-down from weak decays.
The purity and primary fractions of the protons show a sizeable momentum dependence and thus, a \ks-dependent treatment of the $\lambda$ parameters is needed. This is done in three steps. 

First, the purity and primary fraction are determined as functions of \pt. The purity of \sig\ is determined by a \pt-differential fit of the invariant-mass distribution with signal and background templates taken from the simulation. The template fits are shown in Fig.~\ref{fig:invmass_comparison} (left panel) in the relevant \pt\ region. The proton sample is contaminated by mainly four contributions. These are pions, kaons, feed-down from $\Lambda$, and feed-down from \sig. The purity can be determined from MC. The primary fraction is typically determined by DCA template fits. This is, however, not possible due to the machine learning selection and is challenging to do differentially in \pt. Therefore, an alternative approach is used which makes use of the spectra of protons\,\cite{ALICE:2020nkc}, $\Lambda$\,\cite{ALICE:2019avo}, and \sig\,\cite{ALICE:2025hqt} already measured by ALICE. The spectra of the $\Lambda$ and \sig\ are scaled down by the branching ratios of their respective decays branching ratio into protons, weighted by the reconstruction efficiency of the decay proton, and finally projected onto the momentum of the decay proton. The resulting effective spectra of primary and secondary protons can be divided by each other to find the primary fraction as a function of \pt. As a cross-check, the procedure is repeated for kaons and pions\,\cite{ALICE:2020nkc}, which allows an alternative and MC-independent calculation of the purity. The purity determined with data is in good agreement with the simulation, validating the method.
Second, the \pt\ dependence of the purities and primary fraction is projected onto \ks. The projection matrix is calculated from mixed events to improve the statistical precision of the analysis. Finally, the genuine $\lambda$-parameter is calculated by multiplying the purity of the protons, the primary fraction of the protons, and the purity of the \sig.

The genuine fraction of p\mbox{--}\sig\ pairs at vanishing \ks amounts to 79\% and is slightly increasing towards higher \ks. The main contamination is the combinatorial background of the \sig, which contributes 80\% of the total contamination. Another notable contamination is secondary protons paired with signal \sig\ (2.5\%). The total fraction of secondaries in the proton sample is 2.8\% out of which 73\% stem from $\Lambda$ decays. All other contaminations are below 1\%. The contaminations are considered to be independent of \ks in the relevant \ks region.

The phase-space distribution of the p\mbox{--}\sig\ pairs is constructed by event mixing. Forming pairs of particles from different events breaks their correlations while preserving their single-particle momentum distributions, which is desirable for femtoscopy. To avoid any bias in terms of reconstruction efficiency or acceptance, only identified particles from similar events are paired. To this end, the difference in the $z$ coordinate of the reconstructed primary vertex is restricted to $1$~cm and the difference in the total number of tracks at midrapidity ($|y| <0.5$) is restricted to 4.

The correlation function needs to be normalized. The standard approach of event mixing is known to result in a non-femtoscopic rise of the baseline starting at around $\ks = 500$~\MeVc\,\cite{Lisa:2005dd}. This is related to angular correlations of particles in the same event, called minijets. Based on previous analyses, in particular of the p\mbox{--}$\Sigma^0$ correlation function, the slope of this baseline is assumed to vanish towards $\ks=0$\,\cite{ALICE:2018ysd, ALICE:2019buq}. This behavior is also observed in the p\mbox{--}\sig\ correlation function. However, one can note that already below about $\ks = 300$~\MeVc a considerable influence of the strong interaction is expected from model calculations. This makes the normalization difficult. As a compromise, the normalization region is chosen to be $340<k^*<500$~\MeVc~(4 bins). This interval is varied in both directions by $\pm$40~\MeVc (1 bin) to estimate the systematic uncertainty. A constant function (0-th order polynomial) is used to fit the correlation function in the normalization region.

In this article, an alternative approach is presented, called fixed-angle mixing. Since comprehensive studies using this method are still ongoing, the standard event-mixing approach is kept as the default method and the fixed-angle mixing is used as a systematic variation. In the fixed-angle mixing, ten background p\mbox{--}\sig\ pairs are created from each particle pair in the same event. The opening angle between the particles remains the same as for the same-event pair, and the momenta of the particles are randomly drawn from the measured momentum distributions. These distributions are stored in intervals of event multiplicity with a bin size of 4. The resulting correlation function agrees well with the standard method at low \ks, while it shows no rise at higher \ks. This method is attractive because it is self-normalized and eliminates the need for background fits, thus potentially reducing the uncertainty of the measurement. 

The systematic uncertainty is evaluated by making variations to the particle selection, background construction method, and the background fitting. For the variation of the particle selections, the BDT cut-offs of the \sig\ and proton selection are varied by $\pm10$\% each. For the background, the two different background construction methods are considered and the fitting range is varied in the given range. The former variations are done in all possible combinations to account for possible correlation. Assuming a flat distribution, the maximum variation of the data points is divided by $\sqrt{12}$. The uncertainties arising from particle selection and background fitting are similar at low \ks, while the uncertainty from particle selection vanishes towards higher \ks. For $\ks>250$~\MeVc, the systematic uncertainty is dominated by the uncertainty of the background fitting. In general, the uncertainty of the measured correlation function is dominated by the limited statistics.

\section{Results}

\begin{figure}[!htb]
\centering
\includegraphics[width=0.49\textwidth]{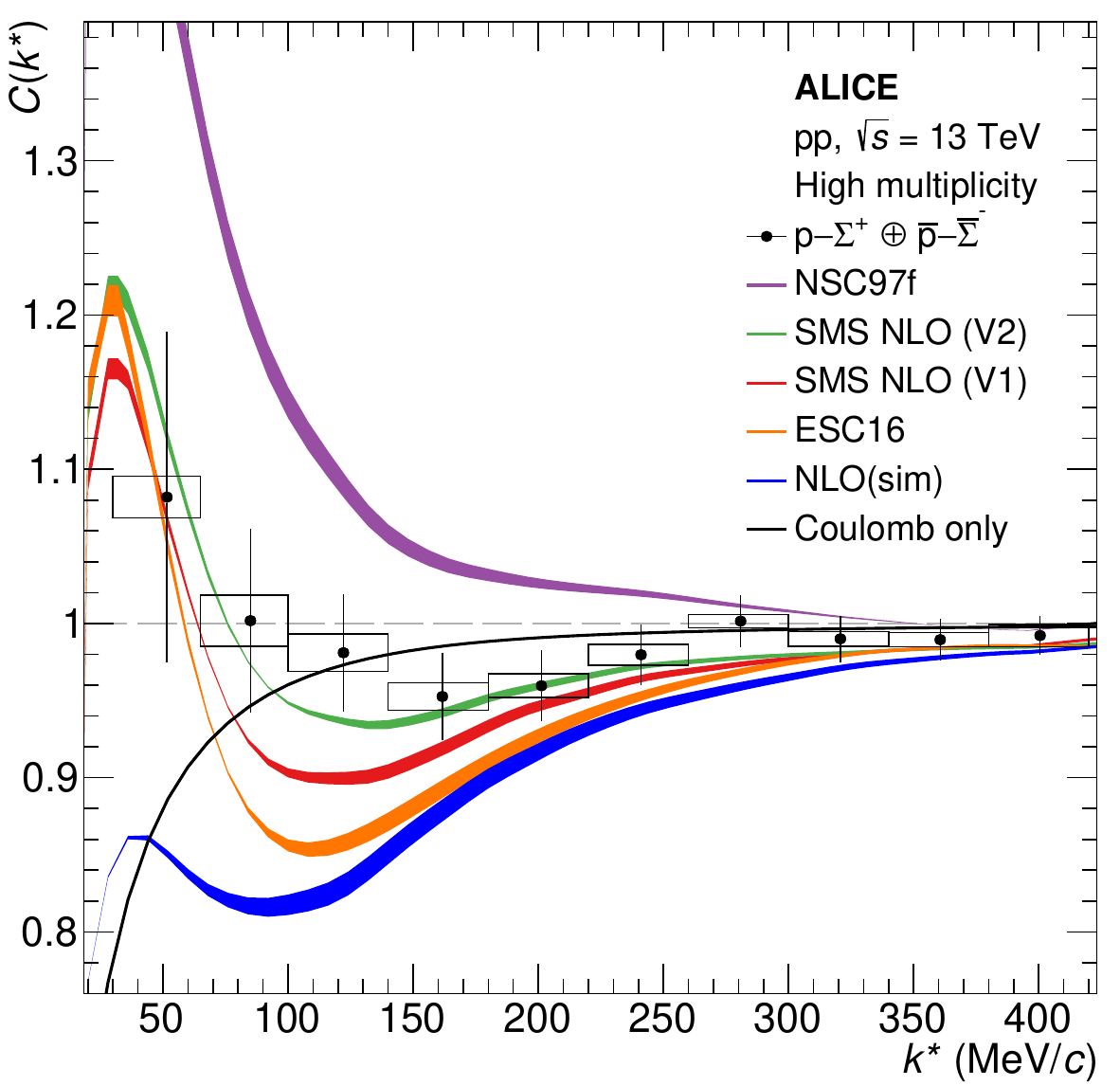}
\includegraphics[width=0.49\textwidth]{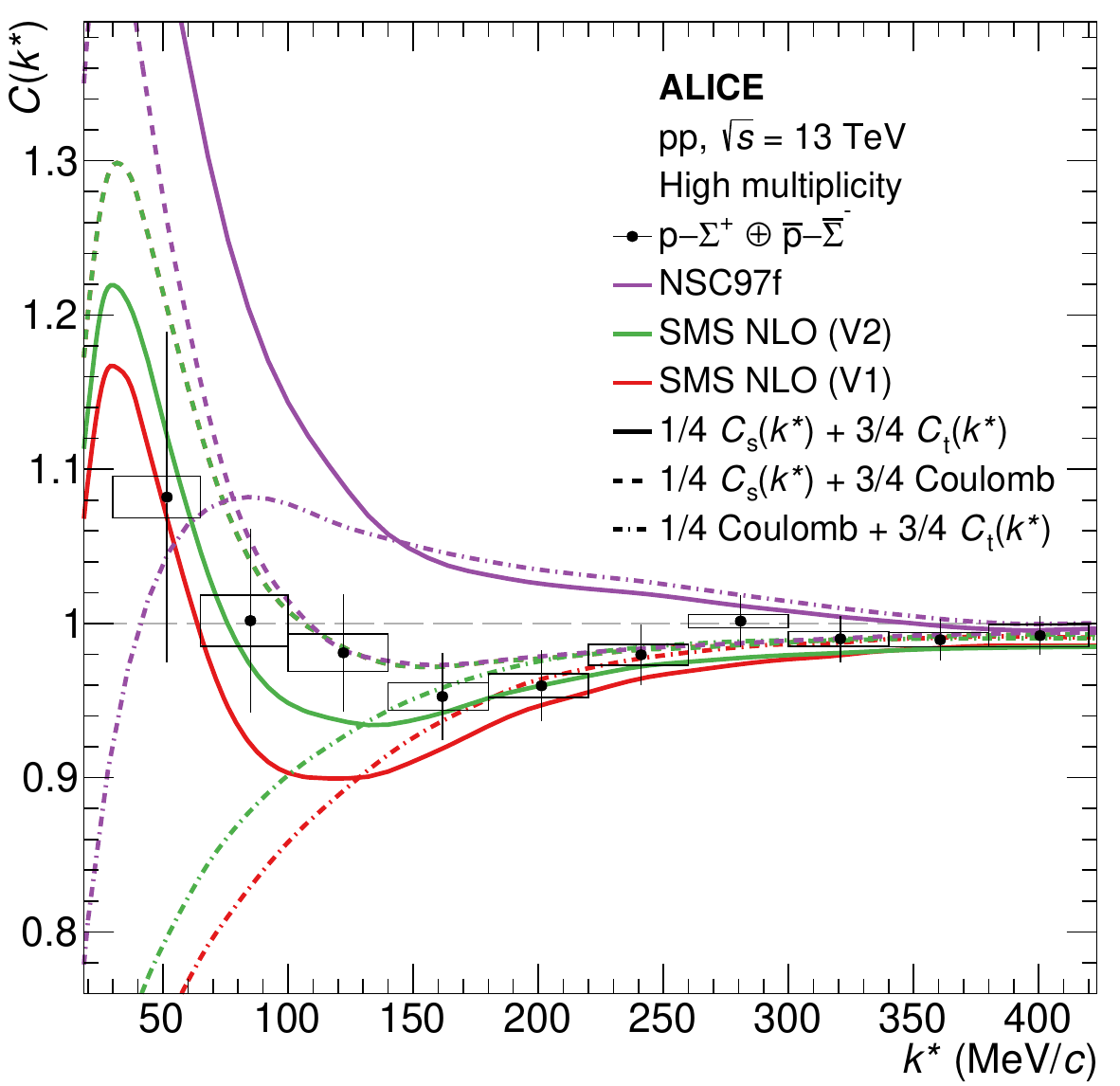}
\caption{p\mbox{--}$\Sigma^+$ correlation function in high-multiplicity triggered pp collisions at \boldmath{$\sqrt{s}=13$}~TeV. The statistical uncertainties are drawn as bars and the systematic ones as boxes. The data points are shifted to the center of gravity of the mixed-event distribution. The data points show the measurement, containing both the genuine contribution as well as the contributions from feed-down and misidentification. The model calculations are weighted by the genuine $\lambda$ parameter and smeared by the momentum resolution to allow a comparison with data.\\Left: Correlation function with several model calculations using the full wave functions and the effective Gaussian parametrization of the source. The uncertainty bands arise from the uncertainty of the source size.\\Right: Decomposition of the theoretical correlation functions into the singlet (dashed lines) and triplet (dash-dotted lines) contributions.  The contributions are multiplied by their statistical weights and added to the Coulomb-only function scaled by 1 – $j_i$, which illustrates the influence of the given contribution on the total correlation function (solid lines).
}
\label{fig:corrfunc_w_r_uncert}
\end{figure}

In Fig.~\ref{fig:corrfunc_w_r_uncert}, the experimental correlation function is shown in comparison with
model predictions.
A bin width of 40~\MeVc is chosen for the correlation function to optimize resolution and statistical significance. The measurement starts at 30~\MeVc because there are only very few counts at smaller \ks. The width of the first two bins is reduced to 35~\MeVc to resolve the peak region of the femtoscopic signal. 
The data points are shifted to the center of gravity of the bin determined from the mixed-event distribution.
Although the uncertainties are sizeable, 
a deviation from unity caused by final-state interactions is evident.

The theoretical framework for the calculation of the correlation function is based on the Koonin\mbox{--}Pratt equation\,\cite{Pratt:1986cc, Lisa:2005dd}:
\begin{align}
 \Cks~=~\int \mathrm{d^3}\rs S(\rs) |\psi(\rs,\ks)|^{2}
\end{align}
where $\psi(\rs,\ks)$ is the two-particle wave function which encodes the interaction among the particles.
The correlation function $\Cks$ is derived by integration of 
$|\psi(\rs,\ks)|^{2}$ over the source distribution \Srs.

For the calculations shown in Fig.~\ref{fig:corrfunc_w_r_uncert}, the source distribution $S(\rs)$ is determined in a data-driven approach using the common-source model\,\cite{ALICE:2020ibs, ALICE:2023sjd, ALICE:2025aur}. In this model, the particle source is described by a Gaussian core surrounded by a halo due to resonance decays. The size of the Gaussian core scales with the mean transverse mass $\langle m_\textrm{T}\rangle$ of the particle pair which in the present  p\mbox{--}\sig\ analysis is ($1.94\pm0.03$)~\GeVcc at $\ks<200$~\MeVc. 
This corresponds to a core radius of 0.85~fm. For the resonance contribution, a simulation-based procedure is used, which is described in detail in Refs.~\cite{ALICE:2020ibs, ALICE:2023sjd}. For the estimate of the uncertainty, the core radius is varied according to the uncertainty of the parametrization with respect to the uncertainty of $\langle m_\textrm{T}\rangle$. For the resonances, the yields, masses, and lifetimes are each varied by 10\%. Assuming a uniform distribution, the total uncertainty is calculated as the difference between the largest and the smallest source of uncertainty divided by $\sqrt{12}$. The complete source,  including the resonance halo, is fitted with a Gaussian in the range from 0.5~fm to 4.5~fm. The resulting effective source size is $(0.98^{+0.03}_{-0.02})$~fm.

The two-particle wave function is often approximated by an asymptotic ansatz, known as the \Ledn approach\,\cite{Lednicky:1981su}. However, the weak interaction in the triplet channel leads to small scattering lengths and, in turn, to large effective ranges, which, together with the small source size, preclude such a simple approximation. Therefore, the theoretical correlation functions are calculated using the full wave functions derived from a solution of the Schrödinger equation. 

The correlation function in the p\mbox{--}\sig\ system contains
the spin singlet and triplet contributions of the strong interaction and
includes also the Coulomb interaction. The spin singlet and triplet scattering lengths ($a_\textrm{s}, a_\textrm{t}$) and effective ranges ($r_s, r_t$) from selected model calculations are listed in Tab.~\ref{Modelparams}. The sign convention of baryon\mbox{--}baryon scattering is adopted so that negative values of the scattering length correspond to attractive interactions. Note that all these potentials were fitted to the p\mbox{--}\sig\ low-energy cross sections of Eisele et al.\,\cite{Eisele:1971mk}. 

\newcommand\ChangeRT[1]{\noalign{\hrule height #1}}
\begin{table}[!htb]
\centering
\caption{Scattering parameters for the p\mbox{--}\sig\ S-waves including
Coulomb effects,
taken from the literature. The variants V1 and V2 are
explained in the text.}
\begin{tabular}{|c|c|c|c|c|c|}
\hline
Interaction & $a_\textrm{s}$~(fm) & $r_\textrm{s}$~(fm) & $a_\textrm{t}$~(fm) & $r_\textrm{t}$~(fm) & Reference(s)\\
\ChangeRT{2pt}
Nagels73 & $-2.42$ & 3.41 & 0.71 & $-0.78$ & \cite{Nagels:1973rq} \\
\hline
NSC97f & $-4.35$ & 3.16 & -0.25 & 28.9 & \cite{Rijken:1998yy} \\
\hline
ESC16 & $-4.30$ & 3.25 & 0.57 & $-3.11$ & \cite{Nagels:2015lfa, Haidenbauer:2021zvr} \\
\hline
Jülich '04 & $-3.60$ & 3.24 & 0.31 & $-12.20$ & \cite{Haidenbauer:2005zh, Haidenbauer:2019boi} \\
\hline
fss2 & $-2.27$ & 4.68 & 0.83 & $-1.52$ & \cite{Fujiwara:2006yh} \\
\hline
NLO\,(sim) & $-2.39$ & 4.61 & 0.80 & $-1.25$ & \cite{Haidenbauer:2021zvr} \\
\hline
NLO19 & $-3.62$ & 3.50 & 0.47 & $-5.77$ & 
\cite{Haidenbauer:2019boi} \\
\hline \hline
SMS NLO (V1) & $-3.62$ & 3.56 & 0.47 & $-6.51$ & \multirow{2}{*}{adapted from~\cite{Haidenbauer:2019boi, Haidenbauer:2023qhf}} \\
\cline{1-5}
SMS NLO (V2) & $-3.62$ & 3.56 & 0.31 & $-16.4$ &  \\
\hline
\end{tabular}
\label{Modelparams}
\end{table}

Figure~\ref{fig:corrfunc_w_r_uncert} shows only those model predictions from Tab.~\ref{Modelparams} where the full wave functions are available to us. These include the Nijmegen soft-core meson-exchange models NSC97f\,\cite{Rijken:1998yy} and ESC16\,\cite{Nagels:2015lfa, Haidenbauer:2021zvr}.
Instead of the $\chi$EFT potential NLO19 (600)\,\cite{Haidenbauer:2019boi}, 
a chiral N\mbox{--}Y potential is used, which is based on a novel regularization scheme, the so-called semilocal 
momentum-space (SMS) regularization\,\cite{Haidenbauer:2023qhf}. Contrary to NLO19, it ensures a realistic description of the $^3\textrm{S}_1$ channel for large momenta (Sect.~3.1 of Ref.~\cite{Haidenbauer:2023qhf}), and consequently of the large-momentum behavior of the correlation function. 
To obtain equivalent results for low momenta, the inherent low-energy constants~(LECs) were tuned to match the scattering parameters of NLO19. This configuration is named SMS NLO~(V1) in Fig.~\ref{fig:corrfunc_w_r_uncert}. A variant with a reduced triplet strength $a_\textrm{t}\approx0.3$~fm is also considered, and named SMS NLO (V2), which will be discussed later. 
The NLO\,(sim)\,\cite{Haidenbauer:2021zvr} potential is based on NLO19, but the LECs were tuned to match the scattering parameters of fss2\,\cite{Fujiwara:2006yh}. This interaction is used as substitute for fss2, for which the full wave functions are not available to us. 
It allows us to explore the full range of predictions from most repulsive (fss2) to attractive (NSC97f) triplet interactions. 
The pure Coulomb part is computed by integrating the Coulomb wave functions over the source. These wave functions are taken from the GSL library\,\cite{gsllib}. The correlation functions for the singlet and triplet interactions are multiplied by their statistical weights $j_i$ (1/4 for the singlet, 3/4 for the triplet) and then added. The correlation function is weighted by the $\lambda$ parameter to account for feed-down and misidentification. The contributions of the contaminations to the correlation function are hereby considered to be flat in \ks. Finally, the correlation functions are smeared by the momentum resolution using the approach implemented in the CATS framework\,\cite{Mihaylov:2018rva}.

In the right panel of Fig.~\ref{fig:corrfunc_w_r_uncert}, the singlet and triplet contributions are shown separately. The contributions are multiplied by their statistical weights and added to the Coulomb-only function scaled by $1-j_i$. In other words, the curves show correlation functions where the singlet or triplet contributions are excluded, to illustrate the actual influence of a given contribution on the total correlation function. In general, the singlet contributions (dashed lines) are  slightly above the data, indicating that a weakly repulsive triplet interaction is missing. Since the singlet interaction
is closely related to that in the nucleon\mbox{--}nucleon system via the approximate
SU(3) flavour symmetry\,\cite{Haidenbauer:2013oca}, it is theoretically 
well constrained, at least on a qualitative level, and is expected to be
attractive.

The correlation function is particularly sensitive to the strength of the less constrained triplet interaction in the \ks region around 100~\MeVc. A repulsive triplet interaction leads to a dip which  scales with the strength of the repulsion. 
The depth of the dip below unity is most important to distinguish the 
theoretical predictions. The data clearly rule out a strong repulsion as predicted by the constituent quark model fss (simulated here by NLO\,(sim)). This finding is in agreement with the latest scattering data\,\cite{J-PARCE40:2022nvq}. 
For illustration, SMS NLO (V1) is re-tuned to produce a triplet scattering length of about 0.3~fm (V2), 
see Table~\ref{Modelparams}.
With that interaction, one can describe the data well, as shown in the left panel of Fig.~\ref{fig:corrfunc_w_r_uncert}. In contrast, a moderate attraction in the triplet channel as predicted by NSC97f is clearly disfavored by the correlation function. This is an important result because the scattering data alone does not suffice to distinguish between an attractive or repulsive triplet interaction\,\cite{Haidenbauer:2013oca,J-PARCE40:2022nvq}.

\begin{figure}[!htb]
\centering
\includegraphics[width=0.49\textwidth]{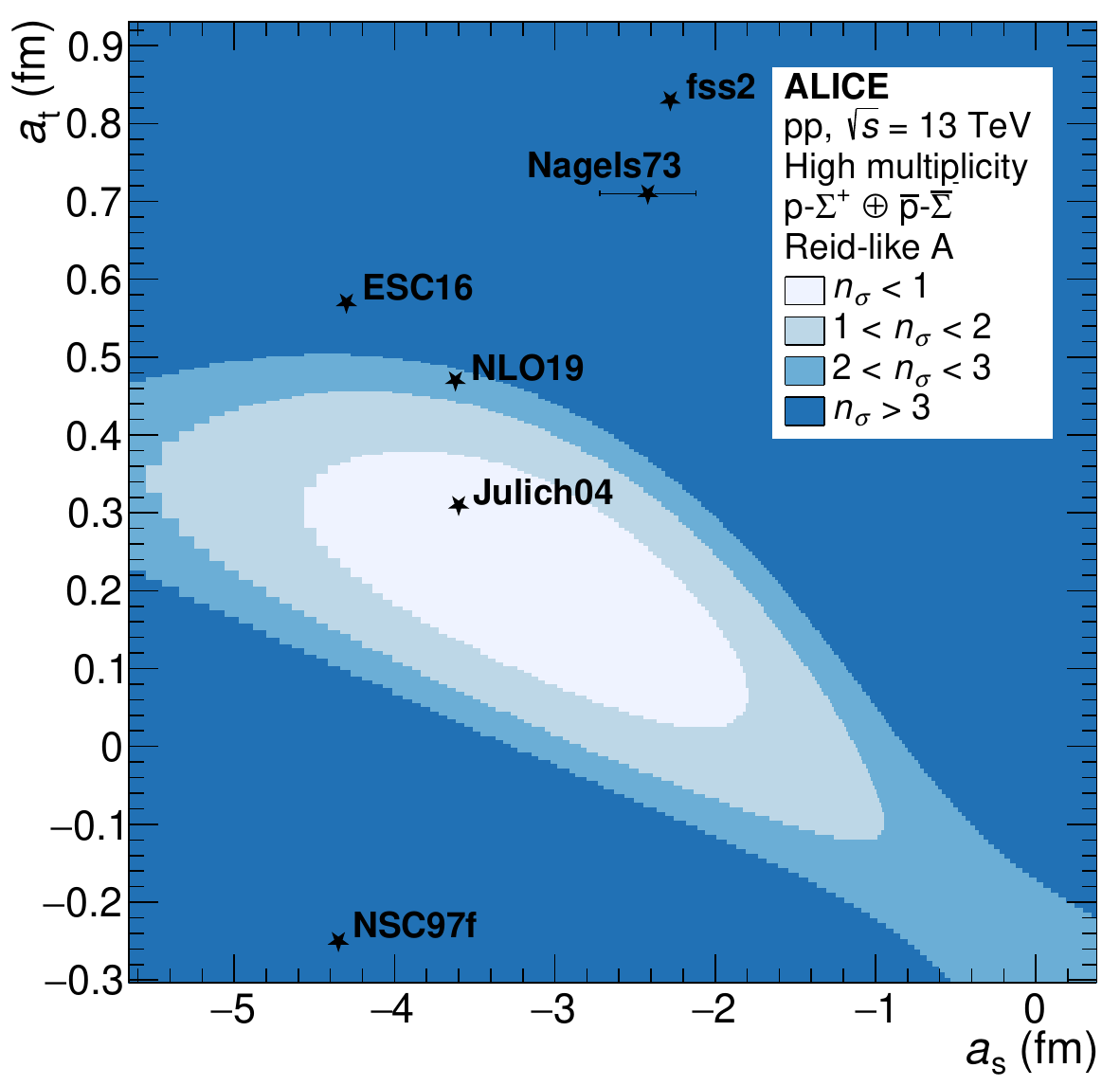}
\includegraphics[width=0.49\textwidth]{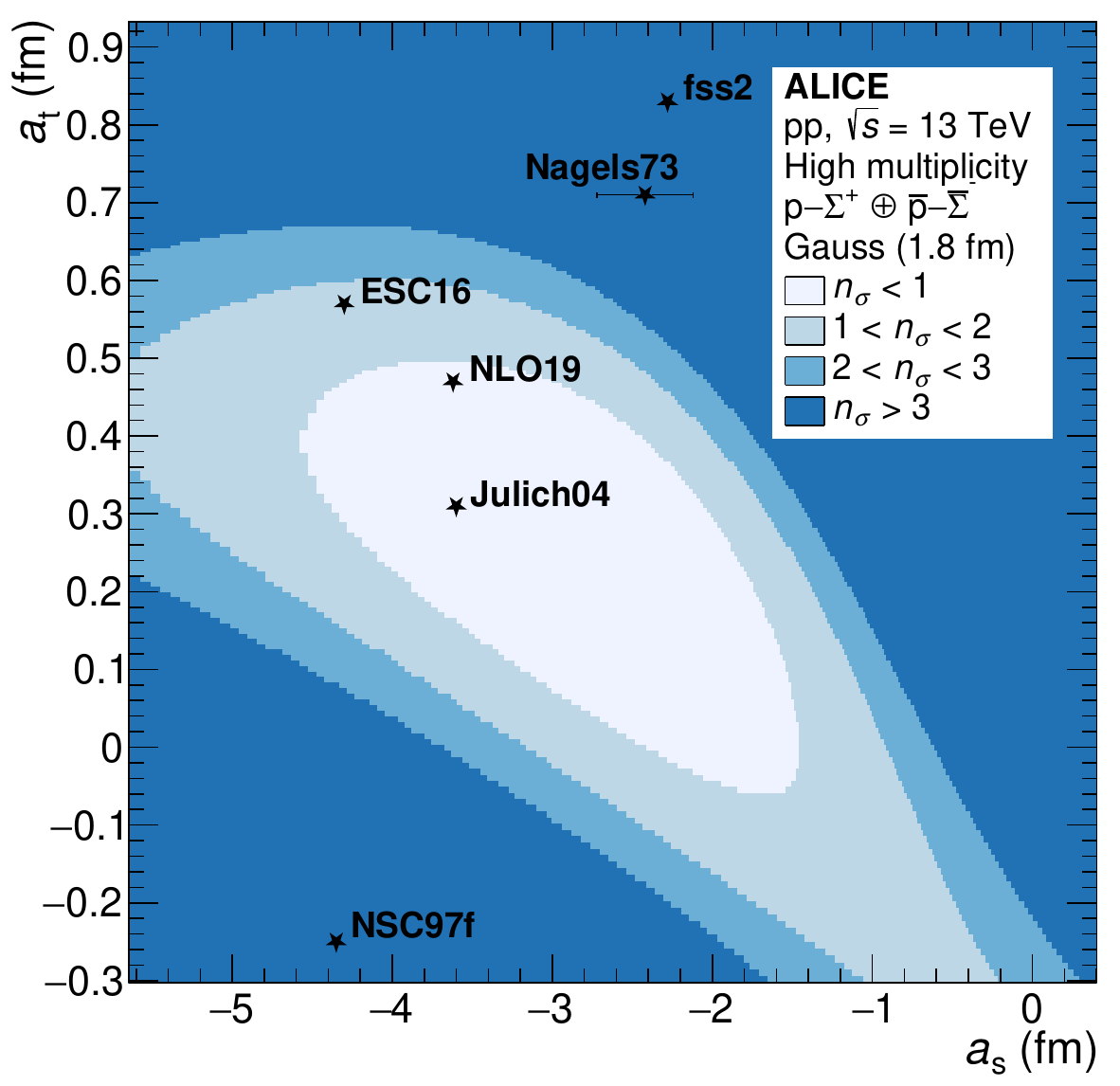}
\caption{Exclusion plots obtained for the Reid-like A (left panel) and Gaussian (1.8~fm) (right panel) potentials as described in the text. The agreement with the data as a function of the singlet and triplet scattering lengths in multiples of the standard deviation is shown together with the model predictions given in Tab.~\ref{Modelparams}.}
\label{excluplot}
\end{figure}

In a statistical analysis of the experimental correlation function, the most probable values for the singlet and triplet scattering lengths were determined. For this purpose, the data are fitted with model correlation functions obtained by solving the Schrödinger equation for a Gaussian potential and for a Reid-like potential. The procedure is as described in Refs.~\cite{Kamiya:2022thy, Morita:2014kza}. As above, for the source function, a Gaussian distribution with radius 0.98~fm was used.
\\For the Gaussian potential, two different choices for the range parameter are considered. Firstly, 1.8~fm is used, in line with values 
found in Ref.~\cite{Gobel:2025afq}, in an attempt to reproduce the low-energy properties of the pp interaction in the $^1\textrm{S}_0$ partial wave with a Gaussian potential. Second, 1.46~fm is employed, corresponding 
to the inverse pion mass, which is naively used as a proxy for the range of pion exchange. In both cases, the potential depths of the singlet and triplet interactions are used as fit parameters.\\For the Reid-like potential, the original Reid NN potential\,\cite{Reid:1968sq} is taken as the starting point. 
It consists of a long-range contribution from pion exchange and a phenomenological short-range part whose parameters were fixed by a fit to the NN phase shifts, as described by Eqs.~(16) and (30) in the given reference. When adapting it for p\mbox{--}$\Sigma^+$, the strength of the pion exchange is re-adjusted in accordance with the SU(3) symmetry\,\cite{Haidenbauer:2013oca}, while 
the strength of the short-range part is varied. This is done by introducing a fit parameter which is, in practice, treated the same way as the depths in the procedure for the Gaussian potentials described above.\\
In the $^1\textrm{S}_0$ channel, the short-range part of the Reid potential 
contains two terms, and two scenarios are considered. 
In scenario A, only the strength of the second (attractive) term is varied, 
while the third (repulsive) term is left unchanged. 
In scenario B, a parametrization is tested, where both short-range terms are modified by a common fit parameter. Evidently, the latter parametrization leads to a potential that is attractive for all values of the fit parameter, which limits the scattering length to be below $-0.9$~fm, the result from pure pion exchange.  This configuration is used as a cross check.\\
Since the interaction in the $^3\textrm{S}_1$ channel is expected to be weak, the coupling to the $^3\textrm{D}_1$ partial wave is neglected, and only the first phenomenological term is taken into account. 
From the given potentials, the S-wave scattering lengths $a_\textrm{s}$ and $a_\textrm{t}$ are calculated. The resulting correlation functions are compared to the data by calculating the total $\chi^2$ and the best-fit function is defined as the one with the smallest $\chi^2$ value. The fit is performed in the range $k^*<180$~\MeVc, above which the simple Gaussian model fails to describe the data. In addition, the correlation functions practically coincide at higher \ks, so no additional information can be extracted by extending the fit range. 
Subsequently, the $\chi^2$ values are related to p-values through a bootstrapping procedure. To this end, $10^5$ artificial correlation functions are generated using the same bin size as the data. In each bin, the value of the best fit function is taken as the central value. This value is replaced by a random value from a Gaussian distribution with a standard deviation equal to the total uncertainty of the respective data point. The resulting correlation functions are fitted again,  resulting in a probability distribution that allows the given $\chi^2$ value of a bin to be related to its p-value. Finally, the p-values are used to define the 1, 2, and 3$\sigma$ intervals using the relation $\textrm{p-value}=\textrm{erf}(n_{\sigma}/\sqrt{2})$. The resulting exclusion plots for the Reid-like and Gaussian potentials are shown in Fig.~\ref{excluplot}, together with several 
model predictions. 

\renewcommand{\arraystretch}{1.5}
\newcommand\ChangeRTv[1]{{\vrule width #1}}
\begin{table}[!htb]
\centering
\caption{Results for the p\mbox{--}\sig\ scattering lengths including
Coulomb effects in the $^1\textrm{S}_0$ ($a_\textrm{s}$) and $^3\textrm{S}_1$ ($a_\textrm{t}$) partial waves, 
obtained by using Reid-like and Gauss potentials for the analysis 
(see text).}
\begin{tabular}{|ccc|ccc|c|}
\hline
Potential  &\ChangeRTv{1.5pt}& Reid-like A               & Gauss (1.8~fm)          &\ChangeRTv{1.5pt}& Reid-like B              & Gauss (1.46~fm)         \\
\hline\hline
$a_\textrm{s}$ (fm) &\ChangeRTv{1.5pt}& $-3.03^{+0.82}_{-1.06}$ & $-2.82^{+0.90}_{-1.20}$ &\ChangeRTv{1.5pt}& $-3.00^{+0.84}_{-1.07}$ & $-2.40^{+0.66}_{-0.78}$ \\
$a_\textrm{t}$ (fm) &\ChangeRTv{1.5pt}& $0.21^{+0.12}_{-0.12}$  & $0.26^{+0.19}_{-0.19}$  &\ChangeRTv{1.5pt}& $ 0.27^{+0.13}_{-0.12}$ & $0.31^{+0.15}_{-0.15}$  \\
\hline
\end{tabular}
\label{tab:fitresults}
\end{table}

The scattering lengths from the statistical analysis and their uncertainties are summarized in Tab.~\ref{tab:fitresults}. The results are in agreement within the given uncertainties, which indicates that there is only a modest model dependence in the analysis. That said, since a Reid-like potential ansatz is physically much better motivated and also more constraining, the values obtained with that interaction are considered to have greater significance. In detail, for the singlet state, attraction is found, where the corresponding potential strength is around 15\% smaller than that for the (Reid) pp potential. In the triplet channel, the data indicate a fairly weak repulsion,  which excludes the two model predictions that produce either a stronger repulsion\,\cite{Fujiwara:2006yh} or an attraction\,\cite{Rijken:1998yy} by more than 3$\sigma$. Interestingly, the scattering parameters of the Jülich~'04 meson-exchange model show the best agreement with the constraints from the measured correlation function and lie within the $1\sigma$ region. The correlation functions corresponding to the $1\sigma$ regions and the correlation function calculated with the Jülich~'04 model are shown in Fig.~\ref{app:supp_fig} in the appendix.

\section{Conclusion}

In this letter, the first measurement of the p\mbox{--}\sig\ femtoscopic correlation function is reported. These data are compared with results based on several p\mbox{--}\sig\ interaction potentials from the literature. 
Given the limited experimental data so far, the theory predictions exhibit significant discrepancies, particularly in the triplet channel. Although the statistical uncertainty of the measured correlation function is still sizeable, the reported measurement makes it possible to distinguish between the model predictions and to exclude some of them by more than 3$\sigma$. In particular, in contrast to scattering experiments\,\cite{Eisele:1971mk}, for the first time, 
one can discriminate between attractive and repulsive triplet interactions. Furthermore, it was possible to obtain a first determination of the p\mbox{--}\sig\ scattering parameters. The data indicate a weak repulsion in the triplet channel, which is best described by the Jülich~'04 meson-exchange model.

Regarding the implications for the possible presence of $\Sigma$ baryons in neutron star matter, sophisticated calculations are required that are beyond the scope of this letter. However, the data presented provide valuable input for such calculations.

This measurement not only provides the most precise constraint on the N\mbox{--}$\Sigma$ interaction to date, but also paves the way for further measurements of the interaction in the p\mbox{--}\sig\ channel, with unprecedented statistics, by ALICE in the LHC Runs 3 and 4.


\newenvironment{acknowledgement}{\relax}{\relax}
\begin{acknowledgement}
\section*{Acknowledgements}
The ALICE Collaboration acknowledges the fruitful discussions with Koji Miwa.

The ALICE Collaboration would like to thank all its engineers and technicians for their invaluable contributions to the construction of the experiment and the CERN accelerator teams for the outstanding performance of the LHC complex.
The ALICE Collaboration gratefully acknowledges the resources and support provided by all Grid centres and the Worldwide LHC Computing Grid (WLCG) collaboration.
The ALICE Collaboration acknowledges the following funding agencies for their support in building and running the ALICE detector:
A. I. Alikhanyan National Science Laboratory (Yerevan Physics Institute) Foundation (ANSL), State Committee of Science and World Federation of Scientists (WFS), Armenia;
Austrian Academy of Sciences, Austrian Science Fund (FWF): [M 2467-N36] and Nationalstiftung f\"{u}r Forschung, Technologie und Entwicklung, Austria;
Ministry of Communications and High Technologies, National Nuclear Research Center, Azerbaijan;
Rede Nacional de Física de Altas Energias (Renafae), Financiadora de Estudos e Projetos (Finep), Funda\c{c}\~{a}o de Amparo \`{a} Pesquisa do Estado de S\~{a}o Paulo (FAPESP) and The Sao Paulo Research Foundation  (FAPESP), Brazil;
Bulgarian Ministry of Education and Science, within the National Roadmap for Research Infrastructures 2020-2027 (object CERN), Bulgaria;
Ministry of Education of China (MOEC) , Ministry of Science \& Technology of China (MSTC) and National Natural Science Foundation of China (NSFC), China;
Ministry of Science and Education and Croatian Science Foundation, Croatia;
Centro de Aplicaciones Tecnol\'{o}gicas y Desarrollo Nuclear (CEADEN), Cubaenerg\'{\i}a, Cuba;
Ministry of Education, Youth and Sports of the Czech Republic, Czech Republic;
The Danish Council for Independent Research | Natural Sciences, the VILLUM FONDEN and Danish National Research Foundation (DNRF), Denmark;
Helsinki Institute of Physics (HIP), Finland;
Commissariat \`{a} l'Energie Atomique (CEA) and Institut National de Physique Nucl\'{e}aire et de Physique des Particules (IN2P3) and Centre National de la Recherche Scientifique (CNRS), France;
Bundesministerium f\"{u}r Forschung, Technologie und Raumfahrt (BMFTR) and GSI Helmholtzzentrum f\"{u}r Schwerionenforschung GmbH, Germany;
General Secretariat for Research and Technology, Ministry of Education, Research and Religions, Greece;
National Research, Development and Innovation Office, Hungary;
Department of Atomic Energy Government of India (DAE), Department of Science and Technology, Government of India (DST), University Grants Commission, Government of India (UGC) and Council of Scientific and Industrial Research (CSIR), India;
National Research and Innovation Agency - BRIN, Indonesia;
Istituto Nazionale di Fisica Nucleare (INFN), Italy;
Japanese Ministry of Education, Culture, Sports, Science and Technology (MEXT) and Japan Society for the Promotion of Science (JSPS) KAKENHI, Japan;
Consejo Nacional de Ciencia (CONACYT) y Tecnolog\'{i}a, through Fondo de Cooperaci\'{o}n Internacional en Ciencia y Tecnolog\'{i}a (FONCICYT) and Direcci\'{o}n General de Asuntos del Personal Academico (DGAPA), Mexico;
Nederlandse Organisatie voor Wetenschappelijk Onderzoek (NWO), Netherlands;
The Research Council of Norway, Norway;
Pontificia Universidad Cat\'{o}lica del Per\'{u}, Peru;
Ministry of Science and Higher Education, National Science Centre and WUT ID-UB, Poland;
Korea Institute of Science and Technology Information and National Research Foundation of Korea (NRF), Republic of Korea;
Ministry of Education and Scientific Research, Institute of Atomic Physics, Ministry of Research and Innovation and Institute of Atomic Physics and Universitatea Nationala de Stiinta si Tehnologie Politehnica Bucuresti, Romania;
Ministerstvo skolstva, vyskumu, vyvoja a mladeze SR, Slovakia;
National Research Foundation of South Africa, South Africa;
Swedish Research Council (VR) and Knut \& Alice Wallenberg Foundation (KAW), Sweden;
European Organization for Nuclear Research, Switzerland;
Suranaree University of Technology (SUT), National Science and Technology Development Agency (NSTDA) and National Science, Research and Innovation Fund (NSRF via PMU-B B05F650021), Thailand;
Turkish Energy, Nuclear and Mineral Research Agency (TENMAK), Turkey;
National Academy of  Sciences of Ukraine, Ukraine;
Science and Technology Facilities Council (STFC), United Kingdom;
National Science Foundation of the United States of America (NSF) and United States Department of Energy, Office of Nuclear Physics (DOE NP), United States of America.
In addition, individual groups or members have received support from:
Czech Science Foundation (grant no. 23-07499S), Czech Republic;
FORTE project, reg.\ no.\ CZ.02.01.01/00/22\_008/0004632, Czech Republic, co-funded by the European Union, Czech Republic;
European Research Council (grant no. 950692), European Union;
Deutsche Forschungs Gemeinschaft (DFG, German Research Foundation) ``Neutrinos and Dark Matter in Astro- and Particle Physics'' (grant no. SFB 1258), Germany;
ICSC - National Research Center for High Performance Computing, Big Data and Quantum Computing and FAIR - Future Artificial Intelligence Research, funded by the NextGenerationEU program (Italy).

\end{acknowledgement}

\bibliographystyle{utphys}   
\bibliography{bibliography}

\newpage
\appendix

\section{Additional figure}

\begin{figure}[!htb]
\centering
\includegraphics[width=0.75\textwidth]{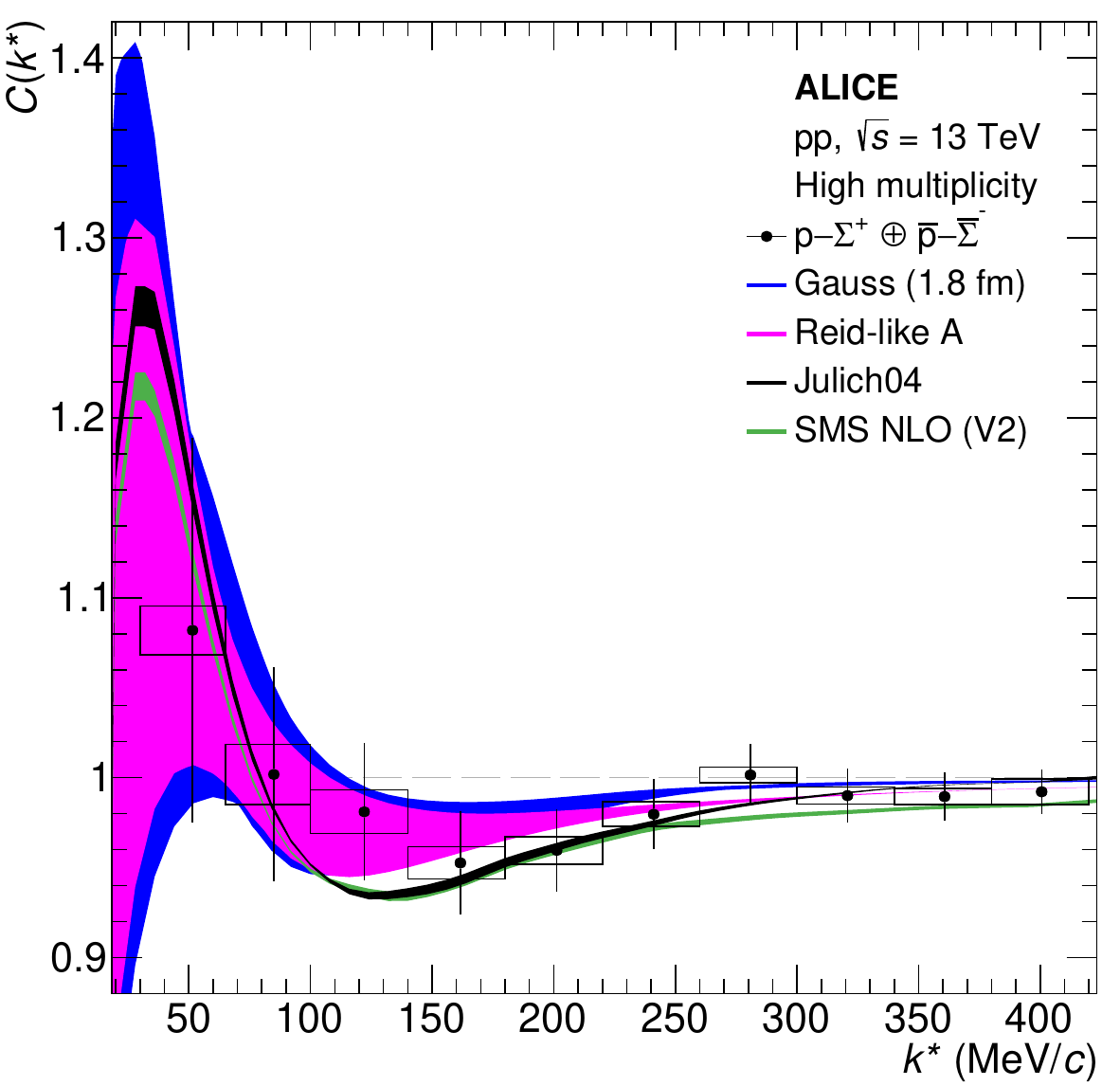}
\caption{p\mbox{--}$\Sigma^+$ correlation function in high-multiplicity triggered pp collisions at $\sqrt{s}=13$~TeV with several model calculations using the full wave functions and the effective Gaussian parametrization of the source. The blue and magenta regions correspond to all parameter sets within the 1$\sigma$ contours of Fig~\ref{excluplot}. The J\"{u}lich~'04 meson-exchange model, which resides within the 1$\sigma$ contours, is shown as a black line. For comparison, the SMS NLO (V2) is also shown as a green line.}
\label{app:supp_fig}
\end{figure}

\clearpage

\section{The ALICE Collaboration}
\label{app:collab}
\begin{flushleft} 
\small

I.J.~Abualrob\,\orcidlink{0009-0005-3519-5631}\,$^{\rm 117}$, 
S.~Acharya\,\orcidlink{0000-0002-9213-5329}\,$^{\rm 50}$, 
G.~Aglieri Rinella\,\orcidlink{0000-0002-9611-3696}\,$^{\rm 32}$, 
L.~Aglietta\,\orcidlink{0009-0003-0763-6802}\,$^{\rm 24}$, 
M.~Agnello\,\orcidlink{0000-0002-0760-5075}\,$^{\rm 29}$, 
N.~Agrawal\,\orcidlink{0000-0003-0348-9836}\,$^{\rm 25}$, 
Z.~Ahammed\,\orcidlink{0000-0001-5241-7412}\,$^{\rm 137}$, 
S.~Ahmad\,\orcidlink{0000-0003-0497-5705}\,$^{\rm 15}$, 
I.~Ahuja\,\orcidlink{0000-0002-4417-1392}\,$^{\rm 36}$, 
Z.~Akbar$^{\rm 83}$
% ...

\section*{Affiliation Notes}

$^{\rm I}$ Deceased\\
$^{\rm II}$ Also at: Max-Planck-Institut fur Physik, Munich, Germany\\
$^{\rm III}$ Also at: Czech Technical University in Prague (CZ)\\
$^{\rm IV}$ Also at: Italian National Agency for New Technologies, Energy and Sustainable Economic Development (ENEA), Bologna, Italy\\
$^{\rm V}$ Also at: Instituto de Fisica da Universidade de Sao Paulo\\
$^{\rm VI}$ Also at: Dipartimento DET del Politecnico di Torino, Turin, Italy\\
$^{\rm VII}$ Also at: Department of Applied Physics, Aligarh Muslim University, Aligarh, India\\
$^{\rm VIII}$ Also at: Institute of Theoretical Physics, University of Wroclaw, Poland\\
$^{\rm IX}$ Also at: Facultad de Ciencias, Universidad Nacional Aut\'{o}noma de M\'{e}xico, Mexico City, Mexico\\
$^{\rm X}$ Also at: Department of Physics, Tohoku University, Japan\\

\section*{Collaboration Institutes}

$^{1}$ A.I. Alikhanyan National Science Laboratory (Yerevan Physics Institute) Foundation, Yerevan, Armenia\\
$^{2}$ AGH University of Krakow, Cracow, Poland\\
$^{3}$ Bogolyubov Institute for Theoretical Physics, National Academy of Sciences of Ukraine, Kyiv, Ukraine\\
$^{4}$ Bose Institute, Department of Physics  and Centre for Astroparticle Physics and Space Science (CAPSS), Kolkata, India\\
$^{5}$ California Polytechnic State University, San Luis Obispo, California, United States\\
$^{6}$ Central China Normal University, Wuhan, China\\
$^{7}$ Centro de Aplicaciones Tecnol\'{o}gicas y Desarrollo Nuclear (CEADEN), Havana, Cuba\\
$^{8}$ Centro de Investigaci\'{o}n y de Estudios Avanzados (CINVESTAV), Mexico City and M\'{e}rida, Mexico\\
$^{9}$ Chicago State University, Chicago, Illinois, United States\\
$^{10}$ China Nuclear Data Center, China Institute of Atomic Energy, Beijing, China\\
$^{11}$ China University of Geosciences, Wuhan, China\\
$^{12}$ Chungbuk National University, Cheongju, Republic of Korea\\
$^{13}$ Comenius University Bratislava, Faculty of Mathematics, Physics and Informatics, Bratislava, Slovak Republic\\
$^{14}$ Creighton University, Omaha, Nebraska, United States\\
$^{15}$ Department of Physics, Aligarh Muslim University, Aligarh, India\\
$^{16}$ Department of Physics, Pusan National University, Pusan, Republic of Korea\\
$^{17}$ Department of Physics, Sejong University, Seoul, Republic of Korea\\
$^{18}$ Department of Physics, University of California, Berkeley, California, United States\\
$^{19}$ Department of Physics, University of Oslo, Oslo, Norway\\
$^{20}$ Department of Physics and Technology, University of Bergen, Bergen, Norway\\
$^{21}$ Dipartimento di Fisica, Universit\`{a} di Pavia, Pavia, Italy\\
$^{22}$ Dipartimento di Fisica dell'Universit\`{a} and Sezione INFN, Cagliari, Italy\\
$^{23}$ Dipartimento di Fisica dell'Universit\`{a} and Sezione INFN, Trieste, Italy\\
$^{24}$ Dipartimento di Fisica dell'Universit\`{a} and Sezione INFN, Turin, Italy\\
$^{25}$ Dipartimento di Fisica e Astronomia dell'Universit\`{a} and Sezione INFN, Bologna, Italy\\
$^{26}$ Dipartimento di Fisica e Astronomia dell'Universit\`{a} and Sezione INFN, Catania, Italy\\
$^{27}$ Dipartimento di Fisica e Astronomia dell'Universit\`{a} and Sezione INFN, Padova, Italy\\
$^{28}$ Dipartimento di Fisica `E.R.~Caianiello' dell'Universit\`{a} and Gruppo Collegato INFN, Salerno, Italy\\
$^{29}$ Dipartimento DISAT del Politecnico and Sezione INFN, Turin, Italy\\
$^{30}$ Dipartimento di Scienze MIFT, Universit\`{a} di Messina, Messina, Italy\\
$^{31}$ Dipartimento Interateneo di Fisica `M.~Merlin' and Sezione INFN, Bari, Italy\\
$^{32}$ European Organization for Nuclear Research (CERN), Geneva, Switzerland\\
$^{33}$ Faculty of Electrical Engineering, Mechanical Engineering and Naval Architecture, University of Split, Split, Croatia\\
$^{34}$ Faculty of Nuclear Sciences and Physical Engineering, Czech Technical University in Prague, Prague, Czech Republic\\
$^{35}$ Faculty of Physics, Sofia University, Sofia, Bulgaria\\
$^{36}$ Faculty of Science, P.J.~\v{S}af\'{a}rik University, Ko\v{s}ice, Slovak Republic\\
$^{37}$ Faculty of Technology, Environmental and Social Sciences, Bergen, Norway\\
$^{38}$ Frankfurt Institute for Advanced Studies, Johann Wolfgang Goethe-Universit\"{a}t Frankfurt, Frankfurt, Germany\\
$^{39}$ Fudan University, Shanghai, China\\
$^{40}$ Gangneung-Wonju National University, Gangneung, Republic of Korea\\
$^{41}$ Gauhati University, Department of Physics, Guwahati, India\\
$^{42}$ Helmholtz-Institut f\"{u}r Strahlen- und Kernphysik, Rheinische Friedrich-Wilhelms-Universit\"{a}t Bonn, Bonn, Germany\\
$^{43}$ Helsinki Institute of Physics (HIP), Helsinki, Finland\\
$^{44}$ High Energy Physics Group,  Universidad Aut\'{o}noma de Puebla, Puebla, Mexico\\
$^{45}$ Horia Hulubei National Institute of Physics and Nuclear Engineering, Bucharest, Romania\\
$^{46}$ HUN-REN Wigner Research Centre for Physics, Budapest, Hungary\\
$^{47}$ Indian Institute of Technology Bombay (IIT), Mumbai, India\\
$^{48}$ Indian Institute of Technology Indore, Indore, India\\
$^{49}$ INFN, Laboratori Nazionali di Frascati, Frascati, Italy\\
$^{50}$ INFN, Sezione di Bari, Bari, Italy\\
$^{51}$ INFN, Sezione di Bologna, Bologna, Italy\\
$^{52}$ INFN, Sezione di Cagliari, Cagliari, Italy\\
$^{53}$ INFN, Sezione di Catania, Catania, Italy\\
$^{54}$ INFN, Sezione di Padova, Padova, Italy\\
$^{55}$ INFN, Sezione di Pavia, Pavia, Italy\\
$^{56}$ INFN, Sezione di Torino, Turin, Italy\\
$^{57}$ INFN, Sezione di Trieste, Trieste, Italy\\
$^{58}$ Inha University, Incheon, Republic of Korea\\
$^{59}$ Institute for Advanced Simulation, Forschungszentrum J\"ulich, J\"ulich, Germany\\
$^{60}$ Institute for Gravitational and Subatomic Physics (GRASP), Utrecht University/Nikhef, Utrecht, Netherlands\\
$^{61}$ Institute of Experimental Physics, Slovak Academy of Sciences, Ko\v{s}ice, Slovak Republic\\
$^{62}$ Institute of Physics, Homi Bhabha National Institute, Bhubaneswar, India\\
$^{63}$ Institute of Physics of the Czech Academy of Sciences, Prague, Czech Republic\\
$^{64}$ Institute of Space Science (ISS), Bucharest, Romania\\
$^{65}$ Institut f\"{u}r Kernphysik, Johann Wolfgang Goethe-Universit\"{a}t Frankfurt, Frankfurt, Germany\\
$^{66}$ Instituto de Ciencias Nucleares, Universidad Nacional Aut\'{o}noma de M\'{e}xico, Mexico City, Mexico\\
$^{67}$ Instituto de F\'{i}sica, Universidade Federal do Rio Grande do Sul (UFRGS), Porto Alegre, Brazil\\
$^{68}$ Instituto de F\'{\i}sica, Universidad Nacional Aut\'{o}noma de M\'{e}xico, Mexico City, Mexico\\
$^{69}$ iThemba LABS, National Research Foundation, Somerset West, South Africa\\
$^{70}$ Jeonbuk National University, Jeonju, Republic of Korea\\
$^{71}$ Johann-Wolfgang-Goethe Universit\"{a}t Frankfurt Institut f\"{u}r Informatik, Fachbereich Informatik und Mathematik, Frankfurt, Germany\\
$^{72}$ Korea Institute of Science and Technology Information, Daejeon, Republic of Korea\\
$^{73}$ Laboratoire de Physique Subatomique et de Cosmologie, Universit\'{e} Grenoble-Alpes, CNRS-IN2P3, Grenoble, France\\
$^{74}$ Lawrence Berkeley National Laboratory, Berkeley, California, United States\\
$^{75}$ Lund University Department of Physics, Division of Particle Physics, Lund, Sweden\\
$^{76}$ Marietta Blau Institute, Vienna, Austria\\
$^{77}$ Nagasaki Institute of Applied Science, Nagasaki, Japan\\
$^{78}$ Nara Women{'}s University (NWU), Nara, Japan\\
$^{79}$ National and Kapodistrian University of Athens, School of Science, Department of Physics , Athens, Greece\\
$^{80}$ National Centre for Nuclear Research, Warsaw, Poland\\
$^{81}$ National Institute of Science Education and Research, Homi Bhabha National Institute, Jatni, India\\
$^{82}$ National Nuclear Research Center, Baku, Azerbaijan\\
$^{83}$ National Research and Innovation Agency - BRIN, Jakarta, Indonesia\\
$^{84}$ Niels Bohr Institute, University of Copenhagen, Copenhagen, Denmark\\
$^{85}$ Nikhef, National institute for subatomic physics, Amsterdam, Netherlands\\
$^{86}$ Nuclear Physics Group, STFC Daresbury Laboratory, Daresbury, United Kingdom\\
$^{87}$ Nuclear Physics Institute of the Czech Academy of Sciences, Husinec-\v{R}e\v{z}, Czech Republic\\
$^{88}$ Oak Ridge National Laboratory, Oak Ridge, Tennessee, United States\\
$^{89}$ Ohio State University, Columbus, Ohio, United States\\
$^{90}$ Physics department, Faculty of science, University of Zagreb, Zagreb, Croatia\\
$^{91}$ Physics Department, Panjab University, Chandigarh, India\\
$^{92}$ Physics Department, University of Jammu, Jammu, India\\
$^{93}$ Physics Program and International Institute for Sustainability with Knotted Chiral Meta Matter (WPI-SKCM$^{2}$), Hiroshima University, Hiroshima, Japan\\
$^{94}$ Physikalisches Institut, Eberhard-Karls-Universit\"{a}t T\"{u}bingen, T\"{u}bingen, Germany\\
$^{95}$ Physikalisches Institut, Ruprecht-Karls-Universit\"{a}t Heidelberg, Heidelberg, Germany\\
$^{96}$ Physik Department, Technische Universit\"{a}t M\"{u}nchen, Munich, Germany\\
$^{97}$ Politecnico di Bari and Sezione INFN, Bari, Italy\\
$^{98}$ Research Division and ExtreMe Matter Institute EMMI, GSI Helmholtzzentrum f\"ur Schwerionenforschung GmbH, Darmstadt, Germany\\
$^{99}$ RIKEN iTHEMS, Wako, Japan\\
$^{100}$ Saga University, Saga, Japan\\
$^{101}$ Saha Institute of Nuclear Physics, Homi Bhabha National Institute, Kolkata, India\\
$^{102}$ School of Physics and Astronomy, University of Birmingham, Birmingham, United Kingdom\\
$^{103}$ Secci\'{o}n F\'{\i}sica, Departamento de Ciencias, Pontificia Universidad Cat\'{o}lica del Per\'{u}, Lima, Peru\\
$^{104}$ SUBATECH, IMT Atlantique, Nantes Universit\'{e}, CNRS-IN2P3, Nantes, France\\
$^{105}$ Sungkyunkwan University, Suwon City, Republic of Korea\\
$^{106}$ Suranaree University of Technology, Nakhon Ratchasima, Thailand\\
$^{107}$ Technical University of Ko\v{s}ice, Ko\v{s}ice, Slovak Republic\\
$^{108}$ The Henryk Niewodniczanski Institute of Nuclear Physics, Polish Academy of Sciences, Cracow, Poland\\
$^{109}$ The University of Texas at Austin, Austin, Texas, United States\\
$^{110}$ Universidad Aut\'{o}noma de Sinaloa, Culiac\'{a}n, Mexico\\
$^{111}$ Universidade de S\~{a}o Paulo (USP), S\~{a}o Paulo, Brazil\\
$^{112}$ Universidade Estadual de Campinas (UNICAMP), Campinas, Brazil\\
$^{113}$ Universidade Federal do ABC, Santo Andre, Brazil\\
$^{114}$ Universitatea Nationala de Stiinta si Tehnologie Politehnica Bucuresti, Bucharest, Romania\\
$^{115}$ University of Cape Town, Cape Town, South Africa\\
$^{116}$ University of Derby, Derby, United Kingdom\\
$^{117}$ University of Houston, Houston, Texas, United States\\
$^{118}$ University of Jyv\"{a}skyl\"{a}, Jyv\"{a}skyl\"{a}, Finland\\
$^{119}$ University of Kansas, Lawrence, Kansas, United States\\
$^{120}$ University of Liverpool, Liverpool, United Kingdom\\
$^{121}$ University of Science and Technology of China, Hefei, China\\
$^{122}$ University of Silesia in Katowice, Katowice, Poland\\
$^{123}$ University of South-Eastern Norway, Kongsberg, Norway\\
$^{124}$ University of Tennessee, Knoxville, Tennessee, United States\\
$^{125}$ University of the Witwatersrand, Johannesburg, South Africa\\
$^{126}$ University of Tokyo, Tokyo, Japan\\
$^{127}$ University of Tsukuba, Tsukuba, Japan\\
$^{128}$ Universit\"{a}t M\"{u}nster, Institut f\"{u}r Kernphysik, M\"{u}nster, Germany\\
$^{129}$ Universit\'{e} Clermont Auvergne, CNRS/IN2P3, LPC, Clermont-Ferrand, France\\
$^{130}$ Universit\'{e} de Lyon, CNRS/IN2P3, Institut de Physique des 2 Infinis de Lyon, Lyon, France\\
$^{131}$ Universit\'{e} de Strasbourg, CNRS, IPHC UMR 7178, F-67000 Strasbourg, France, Strasbourg, France\\
$^{132}$ Universit\'{e} Paris-Saclay, Centre d'Etudes de Saclay (CEA), IRFU, D\'{e}partment de Physique Nucl\'{e}aire (DPhN), Saclay, France\\
$^{133}$ Universit\'{e}  Paris-Saclay, CNRS/IN2P3, IJCLab, Orsay, France\\
$^{134}$ Universit\`{a} degli Studi di Foggia, Foggia, Italy\\
$^{135}$ Universit\`{a} del Piemonte Orientale, Vercelli, Italy\\
$^{136}$ Universit\`{a} di Brescia, Brescia, Italy\\
$^{137}$ Variable Energy Cyclotron Centre, Homi Bhabha National Institute, Kolkata, India\\
$^{138}$ Warsaw University of Technology, Warsaw, Poland\\
$^{139}$ Wayne State University, Detroit, Michigan, United States\\
$^{140}$ Yale University, New Haven, Connecticut, United States\\
$^{141}$ Yildiz Technical University, Istanbul, Turkey\\
$^{142}$ Yonsei University, Seoul, Republic of Korea\\
$^{143}$ Affiliated with an institute formerly covered by a cooperation agreement with CERN\\
$^{144}$ Affiliated with an international laboratory covered by a cooperation agreement with CERN.\\

\end{flushleft} 
  
\end{document}